\documentclass[onecolumn, 12pt]{IEEEtran}

\usepackage{amsmath}
\usepackage{amssymb}
\usepackage{amsfonts}
\usepackage{graphicx}
\usepackage{epsfig}
\usepackage{subfigure}
\usepackage{psfrag}
\usepackage{cite}
\usepackage{latexsym}
\usepackage{url}
\usepackage{color}
\usepackage{multirow}

\PassOptionsToPackage{bookmarks={false}}{hyperref}

\begin{document}
\title{Surveillance and Intervention of Infrastructure-Free Mobile Communications: A New Wireless Security Paradigm}
\author{Jie Xu, Lingjie Duan, and Rui Zhang
\thanks{J. Xu is with the School of Information Engineering, Guangdong University of Technology (e-mail: jiexu.ustc@gmail.com). He is also with the Engineering Systems and Design Pillar, Singapore University of Technology and Design.}
\thanks{L. Duan is with the Engineering Systems and Design Pillar, Singapore University of Technology and Design (e-mail:~lingjie\_duan@sutd.edu.sg).}
\thanks{R. Zhang is with the Department of Electrical and Computer Engineering, National University of Singapore (e-mail: elezhang@nus.edu.sg). He is also with the Institute for Infocomm Research, A*STAR, Singapore.}
}

\maketitle

\begin{abstract}
Conventional wireless security assumes wireless communications are rightful and aims to protect them against malicious eavesdropping and jamming attacks. However, emerging infrastructure-free mobile communication networks are likely to be illegally used (e.g., by criminals or terrorists) but difficult to be monitored, thus imposing new challenges on the public security. To tackle this issue, this article presents a paradigm shift of wireless security to the surveillance and intervention of infrastructure-free suspicious and malicious wireless communications, by exploiting legitimate eavesdropping and jamming jointly. In particular, {\emph{proactive eavesdropping}} (via jamming) is proposed to intercept and decode information from suspicious communication links for the purpose of inferring their intentions and deciding further measures against them. {\emph{Cognitive jamming}} (via eavesdropping) is also proposed so as to disrupt, disable, and even spoof the targeted malicious wireless communications to achieve various intervention tasks.
\end{abstract}


\newtheorem{definition}{\underline{Definition}}[section]
\newtheorem{fact}{Fact}
\newtheorem{assumption}{Assumption}
\newtheorem{theorem}{\underline{Theorem}}[section]
\newtheorem{lemma}{\underline{Lemma}}[section]
\newtheorem{corollary}{\underline{Corollary}}[section]
\newtheorem{proposition}{\underline{Proposition}}[section]
\newtheorem{example}{\underline{Example}}[section]
\newtheorem{remark}{\underline{Remark}}[section]
\newtheorem{algorithm}{\underline{Algorithm}}[section]
\newcommand{\mv}[1]{\mbox{\boldmath{$ #1 $}}}

\vspace{-0em}

\section{Introduction}

\begin{figure}
\centering
 \epsfxsize=1\linewidth
    \includegraphics[width=13cm]{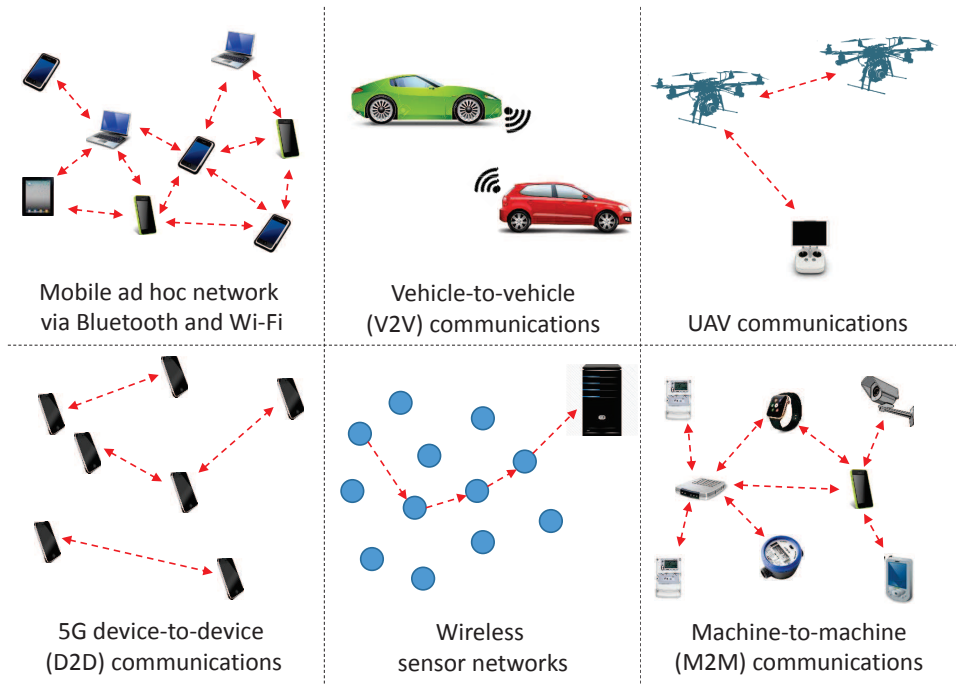}
\caption{Examples of emerging infrastructure-free wireless applications.} \label{fig:1}\vspace{-0em}
\end{figure}

Wireless security has attracted a lot of research interests in the literature. Conventional wireless security studies generally assume wireless communications are rightful, and aim to preserve their confidentiality and availability against malicious eavesdropping and jamming attacks (see, e.g., \cite{ZouWangHanzo2015} and the references therein). However, the presumption of rightful wireless communications may not always hold in practice, especially with the technological advancements in infrastructure-free mobile communications. As shown in Fig. \ref{fig:1}, emerging infrastructure-free wireless applications include mobile ad hoc networks by exploiting Wi-Fi and bluetooth connections,\footnote{For example, FireChat is a mobile chatting software that allows nearby users to interconnect in a mobile ad hoc network by using Wi-Fi and/or Bluetooth locally (see \url{https://www.technologyreview.com/s/525921/the-latest-chat-app-for-iphone-needs-no-internet-connection/}).} unmanned aerial vehicle (UAV) communications, etc. These new infrastructure-free mobile communications can be easily used by malicious users (e.g., criminals, terrorists, and business spies) to commit crimes, jeopardize public safety, invade the secret database of other companies, etc., thus imposing new challenges on the public security. As a response, there is a growing need for authorized parties to surveil and intervene in such mobile communication links, to secure public, commercial, and military benefits. For instance, the National Security Agency (NSA) of the USA has launched the Terrorist Surveillance Program to legitimately monitor wireless devices,{\footnote{{\url{https://nsa.gov1.info/surveillance/}}.}} and the US military uses jammers to protect traveling convoys from cell phone triggered roadside bombs.{\footnote{\url{http://www.methodshop.com/gadgets/reviews/celljammers/}.}}

Facing such new challenges, this article presents a paradigm shift on wireless security from the protection of rightful wireless communications against malicious eavesdropping and jamming (see, e.g., \cite{ZouWangHanzo2015}), to the surveillance and intervention of infrastructure-free suspicious and malicious wireless communications by exploiting legitimate eavesdropping and jamming jointly. In particular, we are interested in the physical-layer design, and focus on a surveillance-and-intervention network consisting of a set of surveillance-and-intervention ``dogs'' (SIDs) and identified suspicious or malicious users, where the SIDs aim to employ the eavesdropping and jamming to achieve two different objectives of surveillance and intervention, respectively.
\begin{itemize}
  \item {\bf Objective I: Surveillance of suspicious communications:}  cooperative SIDs use eavesdropping to intercept and decode the information of suspicious wireless communication links to infer their intentions and decide further measures against them.
  \item {\bf Objective II: Intervention of malicious communications:} SIDs use jamming to disrupt, disable, and even spoof the targeted malicious communications to achieve various intervention tasks.
\end{itemize}

Specifically, we propose a new unified design of eavesdropping and jamming, which are jointly exploited to help each other to improve the surveillance and intervention performances, respectively. In Section \ref{sec:III} we propose proactive eavesdropping on suspicious transmitters via jamming the corresponding receivers, where the SIDs intentionally interfere with the targeted wireless links to alter their communication parameters (such as transmit power, rate, and frequency) to help intercept them more efficiently; and in Section \ref{sec:IV} we propose cognitive jamming to malicious receivers via eavesdropping their corresponding transmitters, where the SIDs eavesdrop targeted wireless links to help design the jamming signals for more efficiently intervening in their communications. To start with, in the following subsections we first provide brief discussions on the deployment issue of SIDs, the identification of suspicious or malicious users, and the related works in the literature, respectively.

\subsection{Using Existing Wireless Infrastructures as SIDs}

The successful surveillance and intervention of suspicious and malicious users require a larger number of densely deployed SIDs to eavesdrop and jam them, as these users are likely to locate anywhere and move from one location to another. Nevertheless, deploying dedicated SIDs is too costly to be implemented in practice. As a result, we propose to use existing wireless infrastructures (such as cellular base stations (BSs) and WiFi access points (APs)) as SIDs for cost-effective information surveillance and intervention.

\begin{figure}
\centering
 \epsfxsize=1\linewidth
    \includegraphics[width=17cm]{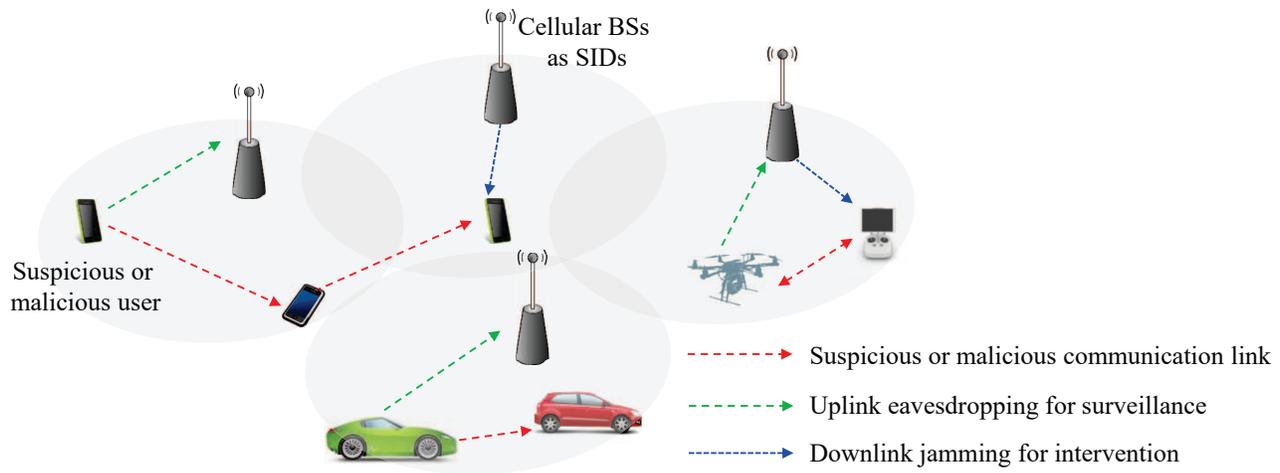}
\caption{Surveillance and intervention of infrastructure-free suspicious and malicious wireless communications via using existing cellular BSs as SIDs.} \label{fig:2}\vspace{-0em}
\end{figure}

Recently, small-cell BSs and WiFi APs have been ultra-densely deployed in cellular networks to provide ubiquitous coverage and high-speed wireless communications; hence, almost any suspicious and malicious users are expected to be within the coverage area of at least one BS/AP. In this case, as shown in Fig. \ref{fig:2}, it is feasible for the SID to use the receive and transmit structures at each BS to eavesdrop and jam the nearby suspicious and malicious users within its coverage area. To successfully implement this, authorized parties need to sign contracts with infrastructure owners and install proper software at the BSs, so as to add the new eavesdropping and jamming functions. The eavesdropping and jamming of SIDs are reminiscent of the uplink and downlink transmissions of cellular BSs, respectively. However, they are different from each other, in the fact that the purposes of eavesdropping and jamming are to overhear and disrupt unintended (suspicious/malicious) users, while uplink and downlink transmissions are for the purpose of communicating with intended users.

It is worth noting that some specific infrastructure-free suspicious and malicious wireless communications are coordinated by existing cellular infrastructures (such as cellular device-to-device (D2D) communications coordinated by BSs \cite{Nardini2016}). In this case, the cellular infrastructure systems can cooperate in sharing information with SIDs, thus helping the surveillance and intervention.



%

\subsection{Identification of Suspicious and Malicious Users}

The successful surveillance and intervention of infrastructure-free wireless communications also require authorized parties to {\it a-priori} identify suspicious and malicious users. In practice, this can be implemented via the big-data analytics (see, e.g., \cite{BiZhangDingCui2015}) based on the massive data collected from communication networks. Though the identification of suspicious and malicious users is not the main focus of this article, we provide brief discussions on several possible methods in the following.
\begin{itemize}
\item {\bf Sensitive Contents Identification}: Malicious users are likely to exchange sensitive contents (such as unpublished business intelligence, sensitive military objects, and confidential terror information) during their communications. Depending on the data types of the contents (e.g., texts, images, audio, or videos), text mining and multimedia data analysis techniques can be applied to extract knowledge and find sensitive information from such unstructured data, so as to identify sensitive objects \cite{BigDataSurvey}.
\item {\bf User Mobility Profiling}: User mobility profiles are also useful information that can help recognize malicious users. For example, a user may be malicious if he/she suddenly deviates from the prior routine trajectories to visit some sensitive places. User mobility profiles can be obtained based on the global positioning system (GPS) information from users' transmitted data (e.g., in Facebook or location-based services). Based on the collected large user mobility profiles, the abnormal ones can be detected by, e.g., applying the outlying sequence detection \cite{Outlying}.{\footnote{It has been shown in \cite{HuiXiong2016} that in public transit systems, rough user mobility profiles collected based on passengers' large-scale transit records are useful to catch malicious users such as pickpocket suspects; while the user mobility profiles extracted from mobile data here are even more accurate. As a result, the user mobility profiling is practically feasible in suspicious user identification.}}
\item {\bf Social Network Map Analysis}: In addition to data analytics on individual users, it is also desirable to explore hidden social connections between these users. For example, terrorists or criminals are usually teamed up to launch critical attacks, and monitoring their mutual information exchange and underlying social activities is useful to identify suspicious and malicious users. Furthermore, leveraging the confirmed terrorist identities and their hidden social connections can also help efficient identification.
\end{itemize}
These methods can be employed jointly with each other or together with other surveillance systems (e.g., video) to improve the identification accuracy. For example, users with abnormal mobility profiles or with frequent social network contact with identified malicious users can first be identified as possibly malicious, and then we can selectively apply sensitive contents identification and video surveillance analysis on them to further confirm they are malicious or not.

\subsection{Related Works}

In the literature, there have been a handful of methods for eavesdropping infrastructure-based communications (e.g., cellular networks). For example, the NSA has deployed dedicated wiretapping devices in network operators' infrastructures, and the eavesdroppers can install monitoring software such as FlexiSPY{\footnote{See {\url{http://www.flexispy.com/}}} in targeted smartphones. Nevertheless, they cannot be efficiently applied to infrastructure-free mobile communication networks. 
On the other hand, passive eavesdropping, which has been conventionally investigated as malicious attacks in the wireless security literature (see, e.g., \cite{ZouWangHanzo2015}), is applicable to intercept infrastructure-free wireless communications. 
However, existing passive eavesdropping schemes only work well when the eavesdroppers are located close to the targeted transmitters with sufficiently strong received signal strength.

There are also some studies on active jamming attacks to intrude and disrupt targeted wireless receivers. For instance, constant, intermittent, reactive, and adaptive jamming schemes have been proposed under different system setups, in which Gaussian noise is artificially generated as the jamming signals to interfere with the targeted receivers \cite{ZouWangHanzo2015}. To efficiently disrupt the communications, these existing schemes require the received jamming signals to be sufficiently strong at the targeted receivers, and thus the jammers should locate close to the targeted receivers and use large transmit power. If we directly apply these schemes to intervene in malicious wireless communications, such jamming activities can be easily detected by malicious receivers and will also cause a lot of victim (legitimate) receivers in the network. In summary, conventional eavesdropping and jamming approaches are independently investigated as two separate lines of research for different attack purposes, while a systematic study on joint eavesdropping and jamming with their cooperative interplay is still missing, thus motivating our study in this article.



\section{Proactive Eavesdropping of Suspicious Communications}\label{sec:III}

First, we consider the eavesdropping of suspicious communication links for the purpose of information surveillance, where SIDs first overhear and decode suspicious signals in the physical layer, and then extract contents from them at higher layers. In order to focus our study on the physical-layer design, we consider that the higher-layer extraction is always successful as long as the physical-layer data is decoded successfully without error. Note that we assume that if the suspicious data is encrypted, then the SID {\it a-priori} knows the key employed for encryption. For example, when suspicious users need to exchange the trusted key before their data transmission, the SID can overhear and acquire such information during the key exchanging phase. On the other hand, if the devices (such as smartphones like BlackBerry and iPhone) can automatically encrypt transmit data for users with the key pre-installed, the SIDs (managed by government agencies) can {\it legally} acquire such information from the device (such as smartphone) manufacturing company.{\footnote{For example, we notice that Canada police has obtained the BlackBerry¡¯s global decryption key (see {\url{https://news.vice.com/article/exclusive-canada-police-obtained-blackberrys-global-decryption-key-how}}), and FBI has accessed iPhone to help in the analysis for a terror case (see {\url{http://www.nbcnews.com/tech/apple/fbi-accessed-iphone-terror-case-raising-more-questions-about-key-n546826}).}} We refer interested readers to \cite{ZouWangHanzo2015} for more details on higher-layer encryption and decryption.}

In practice, the physical-layer eavesdropping of suspicious communications is challenging due to the possible movements of suspicious users, fluctuations of wireless channels, and their co-channel interference. We will propose {\it proactive eavesdropping} to overcome these fundamental challenges.

\begin{figure}
\centering
 \epsfxsize=1\linewidth
    \includegraphics[width=17cm]{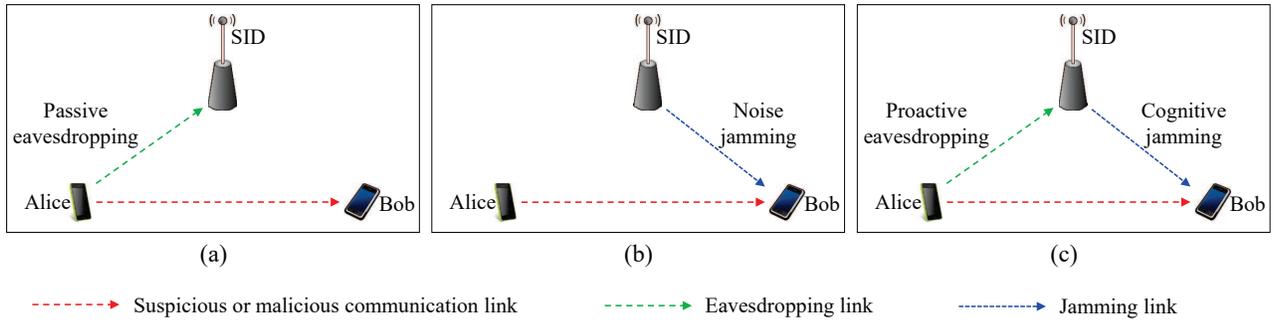}
\caption{Illustration of an SID aiming to eavesdrop or jam a suspicious or malicious wireless communication link from Alice to Bob via (a) conventional passive eavesdropping, (b) conventional jamming, and (c) new proactive eavesdropping and cognitive jamming.} \label{fig:3}\vspace{-0em}
\end{figure}

\subsection{Proactive Eavesdropping}\label{sec:III:A}

To start with, we consider a three-node model as shown in Fig. \ref{fig:3}, where one particular SID aims to eavesdrop a suspicious communication link from Alice to Bob. We introduce the {\it eavesdropping rate} at the physical layer as the fundamental eavesdropping performance metric. Let $R_1$ and $R_0$ denote the achievable data rate of the eavesdropping link from Alice to the SID and the communication rate of the suspicious link from Alice to Bob at the physical layer, respectively. Then, the SID can decode Alice's transmitted signal correctly (with arbitrarily small error) if and only if $R_1$ is no smaller than $R_0$ (i.e., $R_1 \ge R_0$). We define the eavesdropping rate $R_{\rm Eav}$ as the suspicious data rate that the SID can successfully decode, which is given as $R_{\rm Eav} = R_0$ if $R_1 \ge R_0$, and $R_{\rm Eav} = 0$ if $R_1 < R_0$. Our objective is to maximally improve the eavesdropping rate $R_{\rm Eav}$, so as to extract as much information from Alice as possible.

Different from passive eavesdropping in Fig. \ref{fig:3}(a) with given $R_0$ and $R_1$, the SID can use our proposed proactive eavesdropping in Fig. \ref{fig:3}(c) to improve the eavesdropping rate $R_{\rm Eav}$ by purposely changing $R_0$ and/or $R_1$. In proactive eavesdropping, the SID operates in a full-duplex fashion so as to purposely jam and intervene in the suspicious communication to facilitate the simultaneous eavesdropping.\footnote{This approach still works with separately deployed jammer and eavesdropper who are cooperative to enable proactive eavesdropping via jamming.}  Notice that in this approach, the jamming of the SID is implemented at the same time-frequency slots of the eavesdropping, thus introducing self-interference from the jamming to the eavesdropping antenna. In this case, for maintaining the performance of the proactive eavesdropping, the SID should separate its jamming and eavesdropping antennas in distinct locations, and/or employ advanced analog and digital self-interference cancelation methods to minimize the self-interference  \cite{XuDuanZhang2}. Also notice that in practice, it may be difficult to obtain the channel knowledge of the suspicious link as well as the link from the SID to the suspicious receiver. However, it is shown in \cite{XuDuanZhang1} that even knowing these channel distributions (instead of instantaneous gains) is sufficient to exploit proactive eavesdropping for performance enhancement over conventional passive eavesdropping.

\subsubsection{Proactive Eavesdropping via Jamming the Suspicious Receiver}

Depending on the relationship between $R_0$ and $R_1$ (under passive eavesdropping), we consider various proactive eavesdropping approaches by introducing and designing jamming signals to Bob.

{\bf Proactive eavesdropping via jamming:} When $R_1 < R_0$, we have $R_{\rm Eav} = 0$ under the conventional passive eavesdropping. To increase $R_{\rm Eav}$ to be positive in this case, the proactive eavesdropping via jamming approach allows the SID to send jamming signals to interfere with Bob to reduce $R_0$. As long as $R_0$ is reduced to be no larger than $R_1$, then the SID can correctly decode the information signals from Alice, thus improving the eavesdropping rate $R_{\rm Eav}$ to be positive. There are in general two methods to design jamming signals. In the first method, the SID jams Bob with artificially generated Gaussian noise to increase its received noise power, thus reducing the signal-to-noise ratio (SNR) and $R_0$ \cite{XuDuanZhang2,XuDuanZhang1}.  
In the other method, the SID jams Bob by combining Gaussian noise with a processed version of its eavesdropped information signal from Alice \cite{Medard1997,ZengZhang}, where the phase of the eavesdropped signal is carefully designed such that at Bob it is destructively combined with the direct-link signal from Alice. The combined jamming method generally achieves more effective jamming (and thus eavesdropping) performance than the noise jamming method, since it reduces $R_0$ by not only increasing the noise power at Bob's SNR denominator, but also decreasing the useful signal power at the SNR numerator via destructive signal combining. Nevertheless, the combined jamming method is more challenging to be implemented than the noise jamming method, since the former requires the instantaneous knowledge of the channels from both Alice and the SID to Bob, but the latter is applicable even with their partial/statistical knowledge \cite{XuDuanZhang1}. The SID should choose between noise versus combined jamming by considering the eavesdropping performance and complexity tradeoff.


{\bf Proactive eavesdropping via relaying:} When $R_1 > R_0$, we have $R_{\rm Eav} = R_0$ under passive eavesdropping. To further improve $R_{\rm Eav}$ in this case, the proactive eavesdropping via relaying approach allows the SID to operate as a full-duplex relay to receive the signal from Alice and forward to Bob, such that the forwarded signal is constructively combined with the direct-link signal to increase Bob's received SNR and the communication rate $R_0$, thus improving $R_{\rm Eav}$ (as long as the increased $R_0$ is still no larger than $R_1$) \cite{ZengZhang}.  
Note that due to the fading nature of wireless channels, $R_1 < R_0$ and $R_1 > R_0$ may hold over different time instances, and accordingly, the SID needs to adaptively switch between jamming and relaying modes. 

For the purpose of illustration, we provide a numerical example to compare the performance of the proactive eavesdropping versus the conventional passive eavesdropping. In the simulation, we consider that the distance from Alice to Bob is 500 meters, with the x-y coordinates of Alice and Bob being (0, 0) and (0, 500 meters), respectively. We then place the SID at the x-y coordinate of $(0,x)$, where $x \ge 0$ corresponds to the distance from the SID to Alice. For wireless channels, we consider only pathloss by setting the pathloss exponent being 3 and the reference pathloss being $-$60 dB at a distance of 10 meters. Furthermore, we set the transmit powers at both Alice and the SID as 43 dBm, and the receiver noise power as $-$80 dBm. The SID is assumed to be able to perfectly cancel its self-interference from the jamming to the eavesdropping antennas. In this case without any jamming or relaying, the SNR at Bob receiver is around 12 dB and the corresponding communication rate or channel capacity from Alice to Bob is about 4.1 bps/Hz. Under this setup, Fig. \ref{fig:4:new} shows the eavesdropping rate versus the distance from the SID to Alice $x$, and we have the following observations.
\begin{itemize}
  \item When 0 $\le x \le$ 500 meters, it is observed that the passive eavesdropping achieves a constant eavesdropping rate that equals to the communication rate from Alice to Bob, which is due to the fact that the wireless channel from Alice to the SID is stronger than or same as that to Bob. By contrast, the proactive eavesdropping is observed to result in higher eavesdropping rate than the passive eavesdropping in this case, which is achieved via the SID by relaying the information (via amplify-and-forward) from Alice to Bob to increase its received signal strength. In particular, when $x=$ 0, the proactive eavesdropping is observed to achieve a 50\% eavesdropping rate gain (from 4 to 6 bps/Hz) as compared to the passive eavesdropping. The reason is as follows. In this case, the SNR at the SID receiver is sufficiently large and the SID can receive a nearly noise-free signal from Alice. When the SID forwards such a signal with the same transmit power 43 dBm as Alice, 4 dB SNR increase and thus around 2 bps/Hz rate increase at the Bob receiver is achieved with coherent signal combining therein. In addition, when $x$ increases from 0 to 230 meters, the eavesdropping rate gain by the proactive eavesdropping is observed to increase from 50\% to 75\%. This is due to the fact that as $x$ increases, the wireless channel from Alice to the SID receiver worsens (for eavesdropping purpose) but that from the SID transmitter to Bob becomes better (for relaying purpose), and $x=$ 230 meters can optimally balance such a tradeoff and accordingly achieve the highest eavesdropping rate.
  \item When $x > $ 500 meters, the passive eavesdropping is observed to achieve zero eavesdropping rate, since the wireless channel from Alice to the SID is weaker than that to Bob. By contrast, the proactive eavesdropping is observed to result in positive eavesdropping rates when $x$ is between 500 and 1180 meters, which is achieved via the SID by jamming Bob to reduce the communication rate $R_0$ until it is equal to the achievable rate $R_1$ of the eavesdropping link \cite{ZengZhang}.
\end{itemize}
In summary, via adjusting its operations between relaying and jamming, significant performance gains are achieved by proactive eavesdropping over passive eavesdropping in terms of eavesdropping rate.

\begin{figure}
\centering
 \epsfxsize=1\linewidth
    \includegraphics[width=12cm]{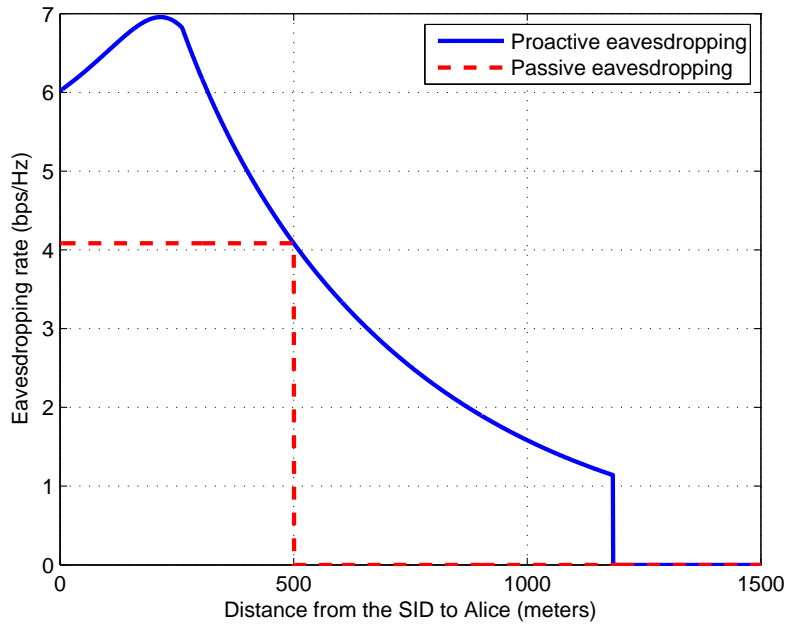}
\caption{The performance comparison between the conventional passive eavesdropping versus the proposed proactive eavesdropping.} \label{fig:4:new}\vspace{-0em}
\end{figure}

\subsubsection{Proactive Eavesdropping via Jamming the Suspicious Transmitter}

Besides jamming Bob, another proactive eavesdropping approach is for the SID to jam and intervene in the suspicious transmitter, namely Alice, by considering the specific communication protocol employed by the suspicious link. Take the time-division-duplex (TDD) multi-antenna transmission scheme as an example, where Alice designs its transmit beamforming vectors based on the reverse-link channel estimation from Bob. In this case, the SID can spoof the reverse-link pilot signals received by Alice, such that Alice estimates a fake channel, and changes its beamforming direction towards the SID and away from Bob (similar to the pilot spoofing attack in \cite{ZhouMahamHjorungnes2012}). This approach increases $R_1$ and decreases $R_0$, and accordingly improves $R_{\rm Eav}$.

The SID can also spoof higher-layer control signals to improve the physical-layer eavesdropping rate $R_{\rm Eav}$. Given $R_1<R_0$, if the suspicious communication employs the hybrid automatic repeat request (HARQ) protocol, then the SID can revise the acknowledgements (ACKs) from Bob to Alice to be negative acknowledgements (NACKs), thus spoofing Alice to increase the number of retransmissions to effectively increase $R_1$ and reduce $R_0$.

\subsubsection{Stealth Eavesdropping}

Since the SID is active in jamming or relaying in the above proposed proactive eavesdropping approaches, it may be detected by Alice and Bob, which can consequently employ advanced anti-eavesdropping methods such as physical-layer security techniques \cite{Kapetanovic2015} and/or anti-jamming methods such as random frequency hopping \cite{XiongLiang}. Therefore, the SID should be sufficiently cognitive to avoid being detected. For example, using Gaussian noise as jamming signals is easier to be detected by Bob via estimating the noise power level. Alternatively, the SID can apply the method of forwarding its eavesdropped signals to destructively combine with the direct-link signal, which reduces the signal power without increasing the noise power. Furthermore, instead of spoofing all ACKs to be NACKs for an HARQ suspicious communication, the SID may selectively spoof only when the direct link is relatively poor in channel quality to reduce its exposure. 

\subsection{Cooperative Eavesdropping}

In practical mobile networks as shown in Fig. \ref{fig:4}, there are possibly many suspicious users or links operating over different frequency bands, and SIDs should cooperate with each other in proactively eavesdropping these links. Depending on the topologies of both the suspicious communication and the SID networks, each SID may operate over multiple bands: some SIDs may jam nearby receivers for the other SIDs' eavesdropping and some SIDs may eavesdrop from others. 
Therefore, it is important to perform a network-wide optimization to determine various SIDs' working modes (e.g., eavesdropping, jamming, or relaying), eavesdropping/jamming frequency bands, and their associations with targeted suspicious users.

Under optimized SID-user associations, different SIDs can use joint detection to cooperatively eavesdrop their common suspicious transmitter target (e.g., SID 1 and SID 2 jointly eavesdrop user 3 in Fig. \ref{fig:4}), or use joint precoding to maximize the jamming power to their common suspicious receiver target (e.g., SID 2 and SID 3 cooperatively jam user 3). Note that both the joint detection (for eavesdropping) and joint precoding (for jamming) for surveillance and intervention are different from those investigated for the BS cooperation in the cellular network such as coordinated multi-point communication (CoMP). The cellular network implements uplink and downlink communications separately over orthogonal time/frequency, whereas the surveillance and intervention need to deal with a mixture of uplink and downlink operations over the same time and frequency slots, thus facing more complicated co-channel interference. As such, the joint eavesdropping and joint jamming for surveillance and intervention entail interesting and yet challenging coordination problems that are worth further pursuing.

\begin{figure}
\centering
 \epsfxsize=1\linewidth
    \includegraphics[width=12cm]{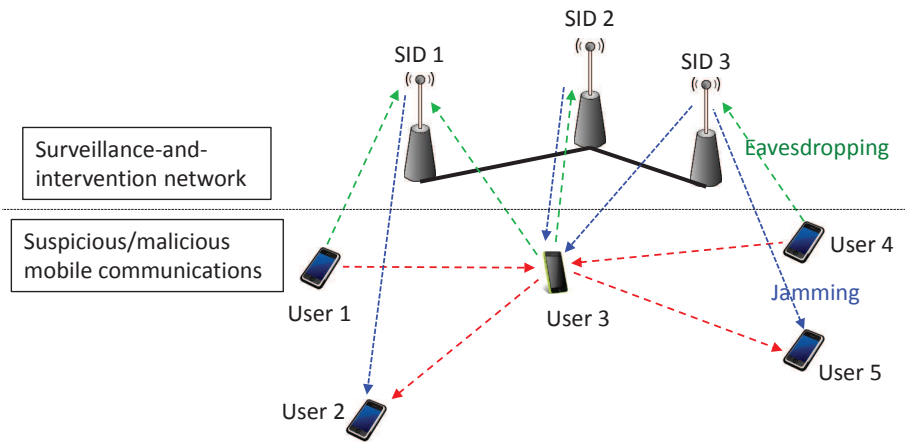}
\caption{An example of multiple SIDs which cooperatively eavesdrop and intervene in multiple suspicious communication links.} \label{fig:4}\vspace{-0em}
\end{figure}

Moreover, in the suspicious mobile network, different users may play different roles or have different importance levels, depending on the network topology and user (node) centrality. For example, user 3 in Fig. \ref{fig:4} is the most critical node, which helps relay the data from user 1 to user 5, as well as that from user 4 to user 2. In this case, as long as the relayed data at user 3 is eavesdropped, the whole network is under successful surveillance. To exploit such properties, SIDs can first coarsely eavesdrop and analyze all suspicious users to determine a priority list of users, and then focus their limited resources on most critical or important nodes.

\section{Cognitive Jamming to Intervene in Malicious Communications}\label{sec:IV}

In this section, we focus on the intervention of infrastructure-free malicious communications for different security purposes. For example, SIDs can actively jam or spoof these malicious communications to interrupt or neutralize the malicious users' plot. In particular, we focus on the physical-layer intervention design, which serves as the basis of any possible higher-layer approaches.


\subsection{Cognitive Jamming and Intervention}

We first consider a particular SID aiming to jam or intervene in a malicious communication link from Alice to Bob, as shown in Fig. \ref{fig:3}. The proposed cognitive jamming or intervention approaches can be classified according to different purposes, including disruption, disabling and spoofing, respectively.

\subsubsection{Disrupting and Disabling Malicious Communications}

In this case, the SID sends jamming signals to disrupt or disable malicious communications. One widely adopted method is for the SID to use all its transmission power and send artificially generated Gaussian noise to maximally degrade the malicious communication rate, as shown in Fig. \ref{fig:3}(b). If the rate is reduced below Bob's quality of service (QoS) requirements, then Alice will stop transmission. 
However, the noise jamming is susceptible to detect by Alice and Bob via measuring the noise level. Differently, we propose that the SID can jam by combining the Gaussian noise with a processed version of the eavesdropped information signal from Alice. If such a forwarded signal is sufficiently strong and destructively combined at Bob with the direct-link signal from Alice, then the direct-link signal will be completely canceled without increasing the noise power level. In addition, the SID can spoof higher-layer control signals to disrupt or disable malicious communications. For example, if the malicious communication employs the HARQ protocol, then the SID can spoof the ACKs (from Bob to Alice) to be NACKs, such that Alice will continuously send the same packet or stop transmission due to its detected high packet error rate.

\begin{figure}
\centering
 \epsfxsize=1\linewidth
    \includegraphics[width=12cm]{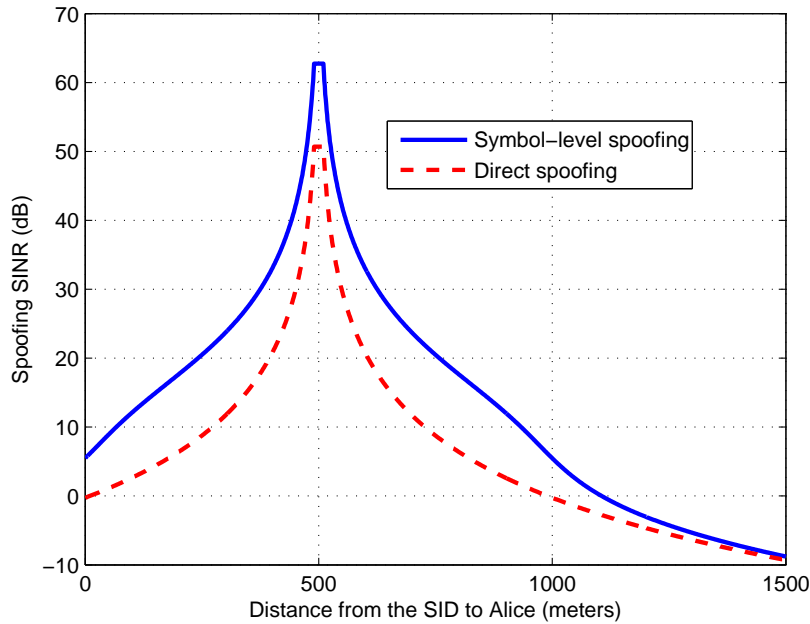}
\caption{The performance comparison between the conventional direct spoofing versus the proposed symbol-level spoofing.} \label{fig:6:new}\vspace{-0em}
\end{figure}

\subsubsection{Spoofing Malicious Communications}

In some other cases, the SID may want to spoof the malicious communication and purposely change the data content received by Bob. For example, the SID can spoof control signals from a malicious user to a UAV and control the UAV to land at a target place to be physically captured. The spoofing of malicious communications is more difficult than the  jamming disruption, since in this case the SID needs to first learn (via eavesdropping) the signal format and the communication protocol employed by the malicious link in order to replace it. After learning, one spoofing approach is for the SID to directly transmit fake signal (with the same format as Alice but including distinct contents), such that Bob will treat the fake signal as the desirable one but the direct-link signal from Alice as noise. However, this requires the received spoofing signal power to be sufficiently larger than that of the direct-link signal and can be easily detected by Bob. Alternatively, the SID can employ a {\it symbol-level} spoofing by sending the superposition of the fake signal and a processed version of the eavesdropped information signal to Bob \cite{XuDuanZhangSpoofing}. With proper signal preprocessing at the SID, the processed signal can offset the direct-link signal and Bob will only receive the fake signal inferring totally different messages. Eavesdropping the transmitted signals from Alice is the key to realize the proposed symbol-level spoofing scheme.

For the purpose of illustration, we provide a numerical example to compare the performances of the direct spoofing method versus the symbol-level spoofing \cite{XuDuanZhangSpoofing}. We use the spoofing signal-to-to-interference-plus-noise ratio (SINR) at Bob as the performance metric, which is defined as the ratio of its received power of the fake signal (from the SID) to that of the direct-link signal (from Alice) plus the noise. By considering the same system setup as in Fig. \ref{fig:4:new}, Fig. \ref{fig:6:new} shows the spoofing SINR versus the distance from the SID to Alice $x$. First, it is observed that the maximum spoofing SINR for both schemes is achieved when the SID is located most close to Bob with $x = 500$ meters. Next, when $0 \le x \le 1000$ meters, more than 6 dB spoofing SINR increase is observed for symbol-level spoofing over the direct spoofing. Furthermore, when the SID is located further away than Alice (with $x > 1000$ meters), the spoofing SINR by the direct spoofing is less than 0 dB. In this case, as the received power of the fake signal (from the SID) is generally weaker than that of the direct-link signal, the direct spoofing cannot be successful. By contrast, the symbol-level spoofing is observed to achieve a spoofing SINR larger than 0 dB when $x \le$  1110 meters. This implies that the symbol-level spoofing can successfully spoof the malicious communication at a much larger distance from the Bob receiver.

\subsection{Cooperative Jamming and Intervention}

Now, we consider the case when multiple SIDs are employed to cooperatively jam or intervene in multiple communication links among malicious users (see an example in Fig. \ref{fig:4}). Since these malicious users may have different intentions, SIDs need to choose the task among disruption, disabling, and spoofing for each of them. Depending on the employed jamming or intervention methods, SIDs may also need to launch eavesdropping from source malicious users to facilitate the disruption or spoofing design. Furthermore, in malicious communication networks, malicious users play different roles in intercommunications, and in this case, as long as the information reception of the most critical malicious users (e.g., user 3 in Fig. \ref{fig:4}) is disabled or spoofed, then the remaining terminals will also be disrupted or spoofed. 

In addition, malicious communications may coexist with many rightful communication links over the same frequency bands. In this case, the jamming and intervention may cause harmful interference to these legal users. Some advanced techniques of cognitive radio \cite{CognitiveRadio} can be applied to protect the communications of rightful users while selectively jamming and intervening in the malicious user communications.

\section{Concluding Remarks}

In this article, we propose a new wireless research paradigm for legitimate mobile communication surveillance and intervention by authorized parties. Unlike conventional wireless security that aims to defend against illegal eavesdropping and jamming, we consider the unique ``reverse'' security perspective to optimize the performances of legitimate eavesdropping and jamming. Due to the space limitation, there are several important issues unaddressed above, which are further discussed in the following.

\subsubsection{Crowdsourcing of Users' Smartphones as SIDs}
In addition to using existing cellular BSs, further crowdsourcing users' smartphones as SIDs is another cost-effective way to realize surveillance and intervention. How to design incentive mechanisms for users to activate their smartphones to track neighboring suspicious and malicious targets is an important issue to be tackled.

\subsubsection{Higher-Layer Encryption and Decryption}

In practice, suspicious users in mobile communication networks may maintain the data confidentiality by using higher-layer cryptographic techniques relying on computational hardness of their underlying mathematical problems and secret key exchange between transmitters and receivers \cite{ZouWangHanzo2015}. 
If such keys are {\it a priori} unknown, then the higher-layer decryption of suspicious communications can become a serious problem for contents extraction at SIDs, even when they have strong computation power. How to resolve this issue should be investigated in future work for the success of information surveillance, no matter for infrastructure-free wireless communications considered here, or infrastructure-based communications studied conventionally.


\subsubsection{Countermeasures of the Proposed Surveillance and Intervention}

In general, conventional wireless security techniques by securing wireless communications can be viewed as countermeasures against mobile communication surveillance and intervention. As two sides of the same coin, the advancement of the latter is expected to push the development of the former, and vice versa. 
Game theory \cite{HanZhu2011} can be a useful tool to model and analyze the strategic interplay between them and understand the evolution of both.

\subsubsection{Surveillance versus Privacy}
Beyond technology, the user privacy issue related to our surveillance solution may be a critical public concern. 
Nevertheless, the proper use of surveillance can indeed help protect rightful users' privacy from malicious users' eavesdropping \cite{Lyon2007}. 
From a technological perspective, we believe that conducting research on advanced (wireless) surveillance techniques is necessary and important for the society, as long as these techniques are appropriately used by governments.

\end{document}